\documentclass[twocolumn,aps,prb,superscriptaddress,amsmath,amssymb,showpacs,floatfix,preprintnumbers]{revtex4-2}
\usepackage[latin9,utf8]{inputenc}
\usepackage{float}
\usepackage{textcomp}
\usepackage{amsmath,amssymb,amscd,hyperref}
\usepackage{graphicx}
\makeatletter
\usepackage{url}
\usepackage{grffile}
\usepackage{bbold}
\usepackage{comment}
\usepackage{dirtytalk}
\usepackage{chngcntr}
\usepackage{tabularx}
\usepackage{multirow}
\usepackage{soul}
\usepackage{cleveref}
\usepackage{braket}
\usepackage{makecell}
\DeclareUnicodeCharacter{2212}{-}
\DeclareUnicodeCharacter{0301}{\hspace{-1ex}\'{ }}

\usepackage{dcolumn}
\usepackage{color}
\DeclareGraphicsExtensions{.png .jpg .pdf}
\usepackage{hyperref}
\hypersetup{
     colorlinks = true,
     linkcolor = blue,
     anchorcolor = blue,
     citecolor = blue,
     filecolor = blue,
     urlcolor = blue
     }
\usepackage{braket}
\usepackage{physics}
\usepackage{array}
\usepackage{booktabs}
\usepackage{soul}

\makeatother

\begin{document}

\title{Reversible and irreversible dynamical topological transitions of magnetic Hopfions}

\author{S. Y. Lu}
\affiliation{School of Materials and Physics, China University of Mining and Technology, Xuzhou 221116, P. R. China}
\affiliation{Key Laboratory of Magnetism and Magnetic Functional Materials of MoE, Lanzhou University, Lanzhou 730000, China}

\author{H. M. Dong}
\email{hmdong@cumt.edu.cn}
\affiliation{School of Materials and Physics, China University of Mining and Technology, Xuzhou 221116, P. R. China}

\author{D. X. Yu}
\email{yudx@lzu.edu.cn}
\affiliation{Key Laboratory of Magnetism and Magnetic Functional Materials of MoE, Lanzhou University, Lanzhou 730000, China}

\author{K. Chang}
\email{kchang@zju.edu.cn}
\affiliation{Center for Quantum Matter, School of Physics, Zhejiang University, Hangzhou 310027, P. R. China}

\date{\today}

\begin{abstract}
Magnetic Hopfions are three-dimensional (3D) topological solitons characterized by a nonzero Hopf invariant and offer a promising platform for 3D spintronics. While researchers have widely studied their static stabilization, their nonlinear dynamics under alternating magnetic (AM) fields remain largely unexplored. Using 3D micromagnetic simulations and analytical mode analysis, we show that an AM field drives two qualitatively distinct dynamical regimes of a confined magnetic Hopfion. In the weak-field regime, resonant excitation of intrinsic Hopfion modes induces a nonlinear instability and irreversibly reconfigures the Hopfion into a toron. In contrast, in the strong-field regime, the system undergoes reversible field-locked topological switching at GHz frequencies, with the magnetization periodically alternating between a topologically trivial ferromagnetic configuration and a Hopfion state. The driving frequency selects the switching pathway: a 2 GHz field drives a breathing pathway associated with the low-frequency collective response, whereas a 40 GHz field produces a nonresonant rotational pathway governed by strong Zeeman-torque-driven precession and field locking. These results identify field amplitude and frequency as independent control knobs and reveal that reversible Hopfion switching can arise either from nonlinear continuation of low-frequency collective motion or from nonresonant high-frequency field locking.
\end{abstract}

\maketitle

Magnetic Hopfions are 3D topological solitons whose spin configurations are classified by the homotopy group $\pi_3(S^2)=\mathbb{Z}$ and characterized by a Hopf invariant $Q_{\rm{H}}$ \cite{rybakov2022a,liuThree2020}. Unlike two-dimensional (2D) skyrmions, Hopfions have linked preimages and intrinsically 3D toroidal spin textures. Their nontrivial topology, nanoscale size, and 3D character make them attractive candidates for high-density spintronic, neuromorphic, and topological information devices \cite{donnelly2017a,katmis2025,saji2023}. Theoretical studies have predicted that chiral magnetic nanostructures with appropriate 3D confinement can stabilize Hopfions \cite{taiHopf2018,liuChiral2018}, and recent experiments have enabled their imaging and controlled nucleation in confined magnetic systems \cite{zhengHopfion2023a,yuReal2023,kentHopfions2021}.

Despite these developments, the dynamical stability of Hopfions remains far less understood than their static properties. 2D magnetic skyrmions exhibit important intrinsic dynamical modes, such as rotation and breathing \cite{dongcpl, Jin_2022}. Existing studies for Hopfions have mainly focused on weak excitations, intrinsic resonances, and current-driven dynamics \cite{sobuck2022a,khod2022,raftrey2021,liuNonlinear2022}. These works have revealed rich internal spectra and nonreciprocal dynamical responses. However, how a Hopfion loses or restores its topology under time-periodic magnetic driving remains an open question. This issue matters because AM fields provide a natural means of high-frequency control, but they can also resonantly excite internal modes and trigger nonlinear instabilities. A central question, therefore, is whether AM fields merely destroy Hopfion topology or can coherently and reversibly control 3D topological states.

We address this question by systematically studying how a confined magnetic Hopfion responds to weak and strong AM fields. We find two sharply distinct dynamical regimes. In the weak-field regime, resonant excitation of intrinsic modes destabilizes the Hopfion and drives an irreversible transition to a toron state, whose mechanism remains unclear, as noted in \cite{raftrey2021,huHop2025,ruiz2025}. Moreover, a magnetic-field-driven transition from Hopfions to magnetic torons occurs around 66 mT and extends up to 15 GHz \cite{raftrey2021}. In the strong-field regime, however, the same type of AM driving produces reversible topological switching between a ferromagnetic (FM) configuration and a Hopfion. Importantly, the microscopic switching pathway depends on frequency. A 2 GHz field drives a breathing-mediated transition connected to the low-frequency collective response of the confined Hopfion, whereas a 40 GHz field produces a rotational transition not attributable to the Hopfion resonances resolved in our weak-field spectrum but instead arising from strong, nonresonant Zeeman-torque-driven precession and phase locking to the external drive. These results show that the field amplitude can control the reversibility of the topological response, while the frequency can select the transition pathway. Compared with recent studies of Hopfion stability and field/current-induced transitions \cite{huHop2025,ruiz2025}, the present work focuses on the AM-field-driven dynamical regime and distinguishes resonant irreversible topology breakdown from nonresonant field-locked reversible switching.

\begin{figure}[t]
\centering
\includegraphics[width=0.9\linewidth]{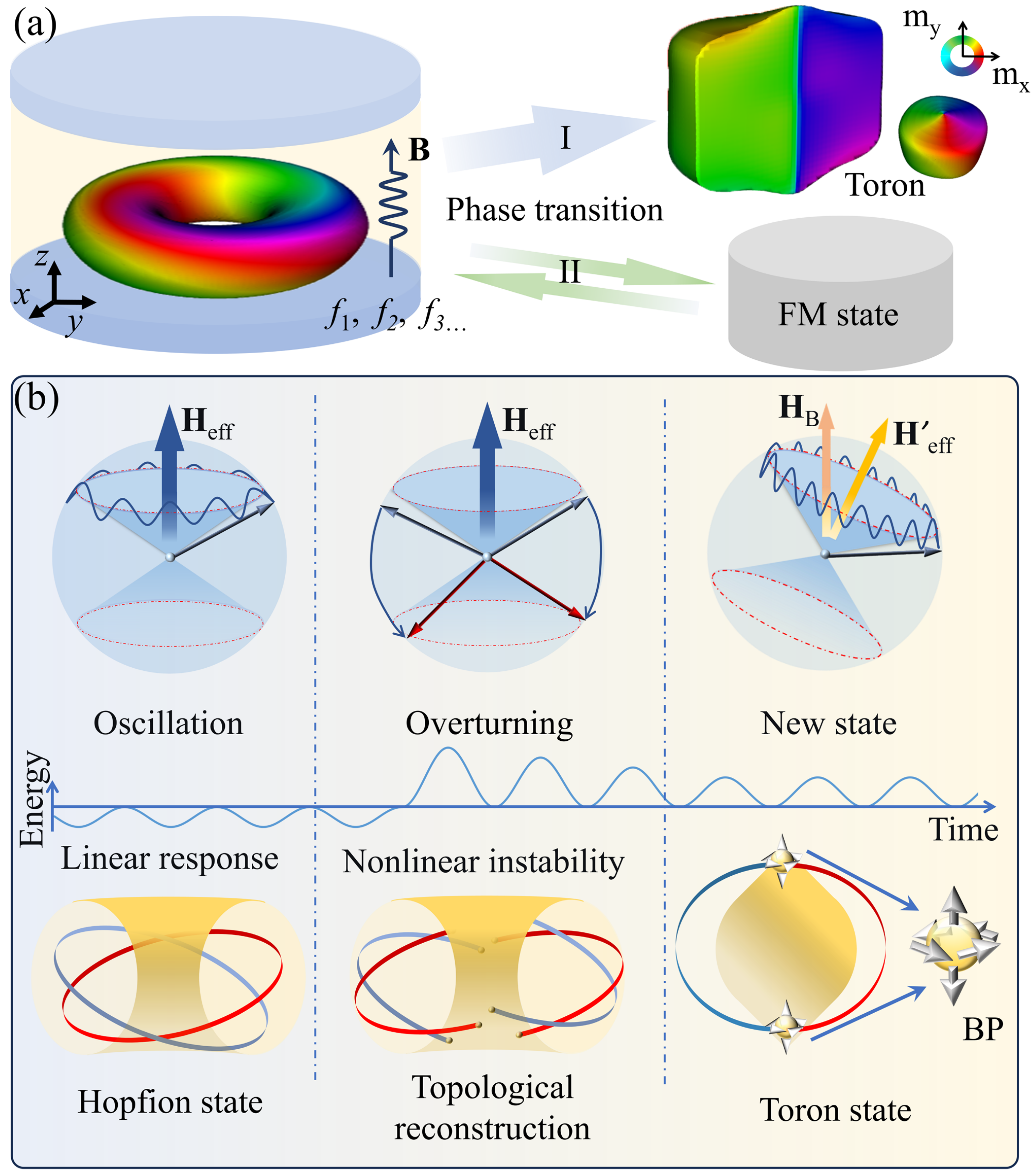}
\caption{(a) Schematic of the cylindrical multilayer nanostructure used to confine a magnetic Hopfion, together with the two dynamical regimes studied in this work. In regime I, weak resonant driving induces an irreversible Hopfion-to-toron transition. In regime II, strong AM driving reversibly switches the system between a topologically trivial FM state and a Hopfion state. (b) Schematic of transition mechanism I, inferred from the time traces of $Q_{\mathrm H}$, energy, and local magnetization dynamics. The energy landscape for the irreversible Hopfion-to-toron reconfiguration and the energy evolution. In the proposed mechanism, resonant nonlinear excitation drives the system over an effective topological barrier, accompanied by local magnetization overturning and the formation of Bloch-point-like singular structures. In the first row, the arrows indicate the instantaneous direction of a magnetic moment, while the cones show its precessional trajectory around the corresponding effective field. The three sketches depict small-angle precession in the linear-response stage, large-angle overturning during the nonlinear instability, and local spin reorientation during topological reconstruction.}
\label{fig-model}
\end{figure}

We consider a cylindrical nanostructure of radius $\mathcal{R}_0=100$ nm and height $\mathcal{H}_0=90$ nm, as shown in Figure~\ref{fig-model}(a). The structure consists of a 70 nm-thick chiral magnetic layer sandwiched between two 10 nm-thick magnetic layers with strong perpendicular magnetic anisotropy (PMA). This geometry provides a 3D confinement potential that stabilizes Hopfions in chiral magnets \cite{kentHopfions2021,liuChiral2018,zhengHopfion2023a}. We obtain the magnetization dynamics by solving the Landau--Lifshitz--Gilbert equation with exchange, Dzyaloshinskii--Moriya interaction (DMI), anisotropy, dipolar interaction, and Zeeman energy terms \cite{liuNonlinear2022}. The micromagnetic model details, material parameters, discretization, and numerical protocol are provided in the Supplemental Material (SM) \cite{supply}.

In the absence of a magnetic field, the confined structure stabilizes a Hopfion with $Q_{\rm{H}}=-1$. 
The normalized Hopf invariant $Q_{\rm{H}}$ is a topological invariant of the 3D magnetization texture, which is calculated by \cite{knapman2025}
\begin{equation}Q_{\rm H}=-\iiint_V d^3r\,
\epsilon_{ijk}A_i F_{jk},\label{eq:Hopf_invariant}
\end{equation}
where $F_{jk}=\mathbf{m}\cdot\left(\partial_j\mathbf{m}\times\partial_k\mathbf{m}\right)$
is the emergent field tensor, and $A_i$ is the corresponding gauge potential satisfying
$F_{jk}=\partial_j A_k-\partial_k A_j$. $\mathbf{m}$ is the normalized magnetization. $i,j,k$ (or equivalently $\{x,y,z\}$) denote the spatial Cartesian coordinates, and repeated indices are summed over according to Einstein's summation convention. The Levi-Civita symbol $\epsilon_{ijk}$ is totally antisymmetric. We can view the texture as a closed, twisted skyrmion tube whose ends join to form a toroidal spin configuration. To probe its dynamics, we use two types of magnetic excitation. First, we apply a weak sinc pulse to probe the intrinsic resonance spectrum. Second, we use sinusoidal AM fields, $B(t)=B_0\sin(2\pi f t)$ with strength $B_0$ and frequency $f$, to investigate nonlinear topological dynamics. As schematically summarized in Fig.~\ref{fig-model}(a), regime I corresponds to the weak-field resonant regime, where the Hopfion undergoes an irreversible Hopfion-to-toron transition, with the proposed nonlinear-instability mechanism illustrated in Figure ~\ref{fig-model}(b), whereas regime II corresponds to the strong-field regime, where reversible FM--Hopfion switching occurs. The mechanism sketched in Fig.~1(b) is a proposed interpretation of the irreversible Hopfion-to-toron transition. The present simulations establish the time evolution of the total energy, $Q_{\rm H}$, and the magnetization texture, but do not spatially resolve the complete topology-changing pathway or directly resolve a Bloch-point core. 

Interpreting $Q_{\rm{H}}$ requires attention to the boundary conditions and the regularity of the magnetization field. Integer-valued Hopf classification applies to a smooth texture whose boundary magnetization permits compactification of the spatial domain. Our finite nanostructure has open boundaries, and large-amplitude driving can produce strong local gradients or allow the texture to unwind at a surface. We therefore assess the topology using the calculated $Q_{\rm{H}}$, the three-dimensional magnetization texture, and, for reversible cycles, the evolution of representative preimages. In particular, $Q_{\rm{H}}$ close to zero alone does not distinguish a toron from a nearly uniform state. Toron identification also relies on its spatial spin structure, whereas Hopfion configurations are characterized by linked preimages. We treat intermediate values during reconstruction as diagnostics of the evolving texture rather than as distinct integer topological sectors.

\begin{figure}[t]
\centering
\includegraphics[width=0.9\linewidth]{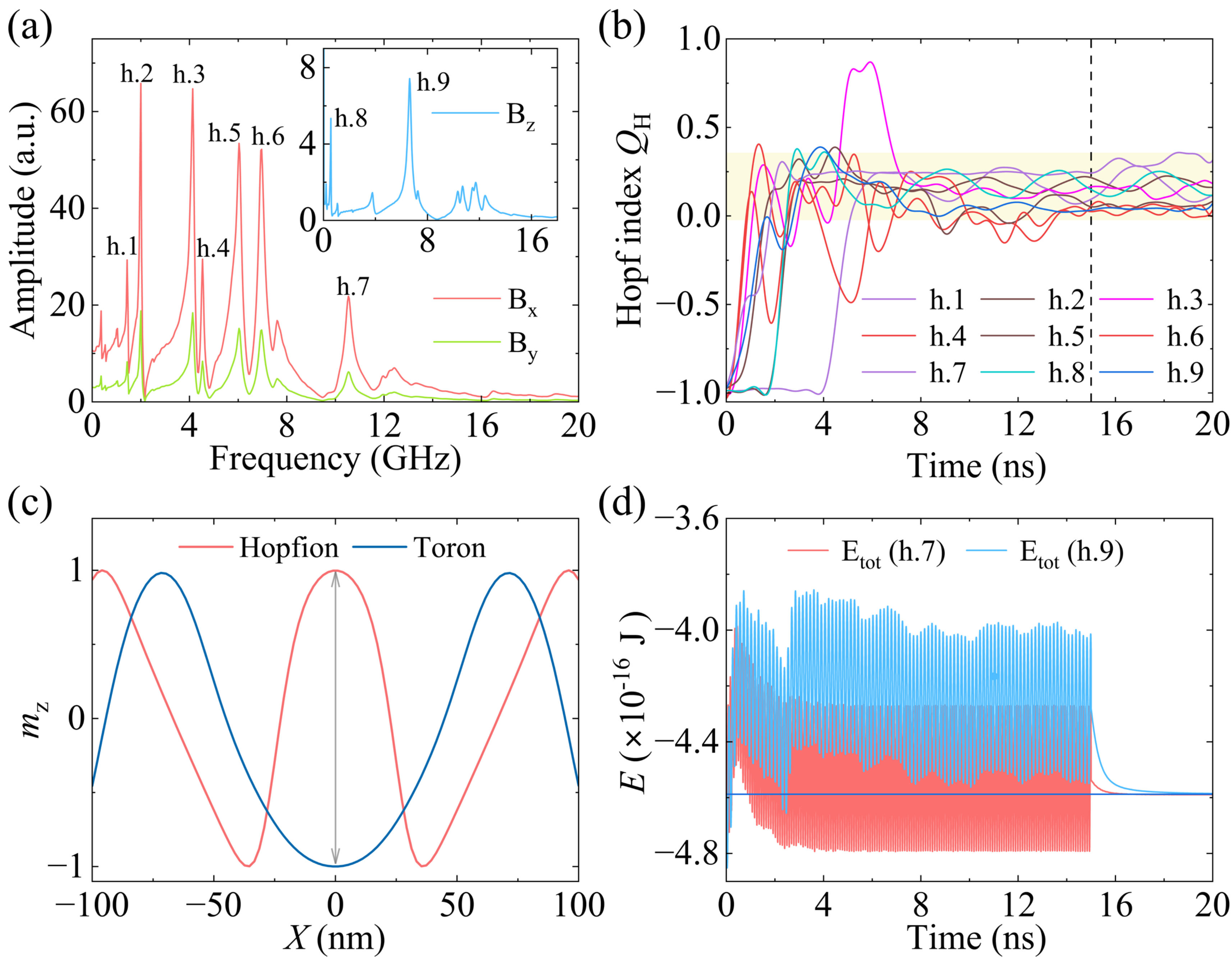}
\caption{(a) Fourier spectrum of the Hopfion response to a weak sinc magnetic-field pulse, revealing intrinsic resonance modes. (b) Time evolution of the Hopf invariant $Q_{\rm{H}}$ under weak sinusoidal AM fields at all resonance frequencies, with $B_0=50$ mT. Resonant driving produces an irreversible Hopfion-to-toron transition. (c) Line profiles of $m_z$ along the Hopfion and toron states' central line through the sample center ($y = z = 0$). The toron state is distinguished by central magnetization reversal. (d) Total energy evolution during the irreversible transition, including the post-drive relaxation interval. The AM field is applied for $0 \leq t < 15$ ns and switched off at $t = 15$ ns. We then follow the dynamics at zero applied field until $t = 20$ ns. The solid blue horizontal line indicates the initial zero-field Hopfion energy.} 
\label{fig-weak}
\end{figure}

We first examine the weak-field regime. Figure~\ref{fig-weak}(a) shows the Fourier amplitude spectrum of the Hopfion response to a sinc pulse $B(t)=B_0\,{\rm sinc}[2\pi f_c(t-t_0)]$, with $B_0=50$ mT, $t_0=5$ ns, and cutoff frequency $f_c=70$ GHz. The response reveals intrinsic Hopfion modes between 0.56 and 10.53 GHz. The strongest resonances occur near 2.00 and 4.1 GHz. The mode amplitudes are well captured by a damped driven-mode description  (see Sec. S2 in the SM \cite{supply}),
\begin{equation}
|A_j|^2=\frac{|\mathcal{F}_j|^2}{(\omega_j-\omega)^2+\Gamma_j^2},
\label{eq:linear}
\end{equation}
where $A_j$ is the amplitude of the $j$th intrinsic mode, $\omega_j=2\pi f_j$ is the eigenfrequency of mode $j$, $\Gamma_j$ is the damping rate, and $\mathcal{F}_j$ denotes the effective field-mode coupling. $\omega =2\pi f$ is the driving frequency. Fig.~S1 in the SM shows the spatial amplitude and phase profiles of the dominant modes in detail. In particular, the in-plane responses are correlated but not identical, consistent with the nonreciprocal dynamics of 3D chiral solitons \cite{liuNonlinear2022}.

When a sinusoidal magnetic field is increased at an intrinsic resonance frequency, the initially stable Hopfion becomes dynamically unstable above a threshold amplitude $B_c^{(j)}$, as shown in Eq. S41 and Figure S2 in the SM \cite{supply}. The detailed theoretical analysis is shown in S2 of the SM \cite{supply}. Figure ~\ref{fig-weak}(b) collects the time evolution of $Q_{\text{H}}$ for all nine resonant drives with $B_0=50$ mT. The field is applied for the first 15 ns and then removed for 5 ns. In each case, the Hopf invariant evolves from its initial value near $-1$ to approximately zero, signaling a transition from a Hopfion to a toron state (see Figure \ref{fig-model}(b)). After the field is switched off, $Q_{\text{H}}$ remains close to zero and the toron texture persists, directly demonstrating the irreversibility of the transition. The transition is irreversible after the field is switched off because a barrier prevents topological reconstruction. The persistence of the toron after field removal is consistent with a barrier
to recovery of the initial Hopfion. The corresponding spin texture contains Bloch-point-like singular structures and differs qualitatively from the original Hopfion. This distinction is also visible in the central line profile of $m_z$ along the $x$ axis through the sample center ($y = z = 0$), shown in Figure~\ref{fig-weak}(c), where the toron state exhibits central magnetization reversal. We identify three dynamical stages from the energy $Q_\text{H}$ and the magnetization texture. In the linear-response stage, the energy exhibits small, nearly periodic oscillations while $Q_\text{H}$ remains near $-1$. The nonlinear-instability stage is marked by strongly modulated energy excursions and large-amplitude magnetization motion. Topological reconstruction is identified by $Q_\text{H} $'s abrupt departure from its initial value, along with localized singular structures or topology-changing deformation at the sample boundary; subsequent energy relaxation indicates entry into low-energy toron states. The energy changes in Figure \ref{fig-weak}(d) and Figure S3 of the SM accompany the texture reconstruction. Because the driven total energy includes a time-dependent Zeeman contribution, these traces do not directly determine a static energy barrier. The toron-like texture persists throughout the additional $5$ ns of zero-field evolution, with no spontaneous recovery of the initial Hopfion. This supports irreversibility on the simulated timescale.

Figure S4 of the SM compares time-aligned energy, $Q_\text{H}$, and local magnetization traces for representative modes h.1 and h.8. At a fixed mesh position near the sample center, small transverse oscillations develop into large-amplitude motion and magnetization reversal. The reversal, the change in $Q_\text{H}$, and the decrease in cycle-minimum energy occur within the same transition interval, supporting the proposed mechanism in Fig. 1(b). Although the modes exhibit different transient energy responses, their magnetization cross sections at 20 ns show toron-like final states (see Fig. S5 of the SM).

The nonlinear instability can be described phenomenologically by a driven nonlinear complex equation for the mode amplitude $A_j$ (see Sec. S2 in the SM \cite{supply}), 
\begin{equation}
i\frac{dA_j}{dt}=(\Delta\omega_j-i\Gamma_j)A_j+\kappa_j |A_j|^2A_j+\mathcal{F}_j,
\label{eq:nonlinear}
\end{equation}
where $\Delta\omega_j=\omega_j-\omega$, $\kappa_j$ is a complex nonlinear coefficient. Equation~\eqref{eq:nonlinear} captures the resonant growth of an internal Hopfion mode and the onset of nonlinear magnetization overturning. Because the system is finite and has open boundaries, the Hopf invariant can change through boundary-mediated deformation and local singular structures, rather than through a smooth topological deformation in an ideal closed continuum. Thus, the weak-field transition represents a resonantly induced topological breakdown of the confined Hopfion.

The amplitude equation describes the weakly nonlinear evolution near an isolated resonance. It separates resonant amplification from subsequent spatial reconstruction: the drive first excites a collective mode, nonlinear frequency shifts and damping modify its response, and sufficiently large deformations can bring the texture to a topology-changing configuration. The critical field is therefore determined from the micromagnetic trajectories, while single-mode theory provides a conditional estimate of its dependence on parameters. In particular, $B_c^{(j)} \propto \alpha$ follows only at effective resonance
and when nonlinear corrections are small, and the reconstruction amplitude and field-mode coupling depend weakly on damping. Detuning, finite excitation duration, and coupling to other modes can invalidate this limiting scaling. We identify the reconstructed state from the magnetization texture and $Q_\text{H}$, and test its persistence by switching off the drive. The toron-like state remains throughout the subsequent 5 ns of zero-field evolution, supporting irreversibility on the simulated timescale.

\begin{figure}[t]
\centering
\includegraphics[width=0.9\linewidth]{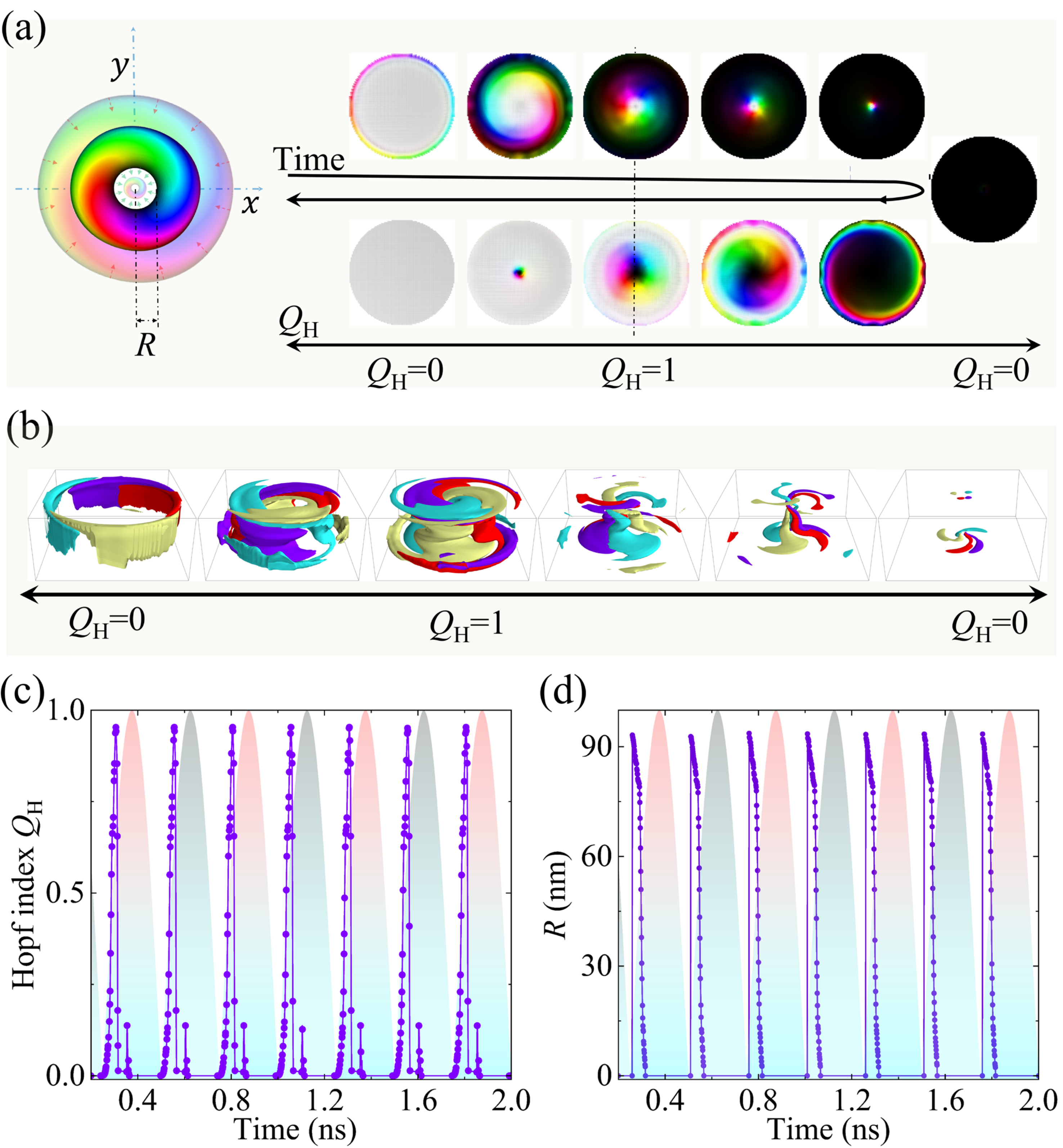}
\caption{(a) Temporal evolution of the magnetization texture during breathing-mediated reversible switching under a strong AM field with $B_0=11$ T and $f=2$ GHz.
(b) Evolution of representative preimage links, showing the periodic emergence and disappearance of Hopfion topology. (c) Time evolution of the Hopf invariant $Q_{\rm{H}}$ synchronized with the applied AM field. (d) Evolution of the characteristic radius $R$ of the central topological structure, demonstrating breathing-type contraction and expansion.}
\label{fig-breathing}
\end{figure}

We next turn to the strong-field regime. Surprisingly, a much stronger AM field does not simply destroy the Hopfion. Instead, it can drive a coherent and reversible topological switching cycle. Figure~\ref{fig-breathing}(a) shows the configuration evolution under a field with $B_0=11$ T and $f=2$ GHz applied along the $z$ direction. Within each switching cycle, the system passes through a nearly FM configuration. From this configuration, the spin texture contracts and develops a linked toroidal structure characteristic of a Hopfion. It then reverts to an FM configuration with reversed magnetization. During the second half-cycle, the process reverses and returns to the original FM state. As shown in Figure~\ref{fig-breathing}(b), the preimage links periodically appear and disappear, directly visualizing the creation and annihilation of Hopfion topology.

The corresponding Hopf invariant oscillates between $Q_{\rm{H}}=0$ and a nonzero value close to 1, as shown in Figure~\ref{fig-breathing}(c). This oscillation is synchronized with the applied field, demonstrating deterministic topological switching. The sign of the Hopf invariant in the dynamically generated Hopfion depends on the orientation of the field-driven winding and may differ from that of the zero-field equilibrium Hopfion. Thus, the dynamically generated Hopfion-like state in the switching cycle is not necessarily identical to the relaxed zero-field Hopfion with $Q_{\rm H}=-1$, including its sign convention. In this strong-field cycle, the transition between the FM and Hopfion states occurs through large-amplitude deformation of the magnetization texture in the finite nanostructure, where boundary effects and transient singular regions enable changes in the Hopf invariant. The 40 GHz rotational switching is therefore not a continuation of a weak-field Hopfion eigenmode, but a strongly driven, nonresonant Floquet-like limit cycle in which the large Zeeman torque phase-locks the azimuthal rotation of the 3D spin texture to the external field.

The switching pathway at 2 GHz is breathing-like. As shown in Fig.~S6(a) of the SM \cite{supply}, after an initial transient of about $0.12$ ns, the system enters a stable reversible topological-switching cycle. To quantify it, here $R$ is extracted from the radius of the contour of the central texture in the $xy$ plane, as shown in Figure~\ref{fig-breathing}(a). Figure~\ref{fig-breathing}(d) shows that $R$ periodically contracts and expands with the same period as the applied field. The total energy of the system, as well as the other free energy components, also undergoes synchronous periodic variations, as shown in Figure S6(b) and (c) of the SM \cite{supply}. The radius $R$ is anticorrelated with the Hopf invariant: Hopfion formation is associated with contraction of the central structure, whereas return to the FM state is associated with expansion and unwinding. The breathing pathway, therefore, reflects strong coupling between the AM field and the low-frequency internal breathing mode of the confined Hopfion. 
\begin{figure}[t]
\centering
\includegraphics[width=0.9\linewidth]{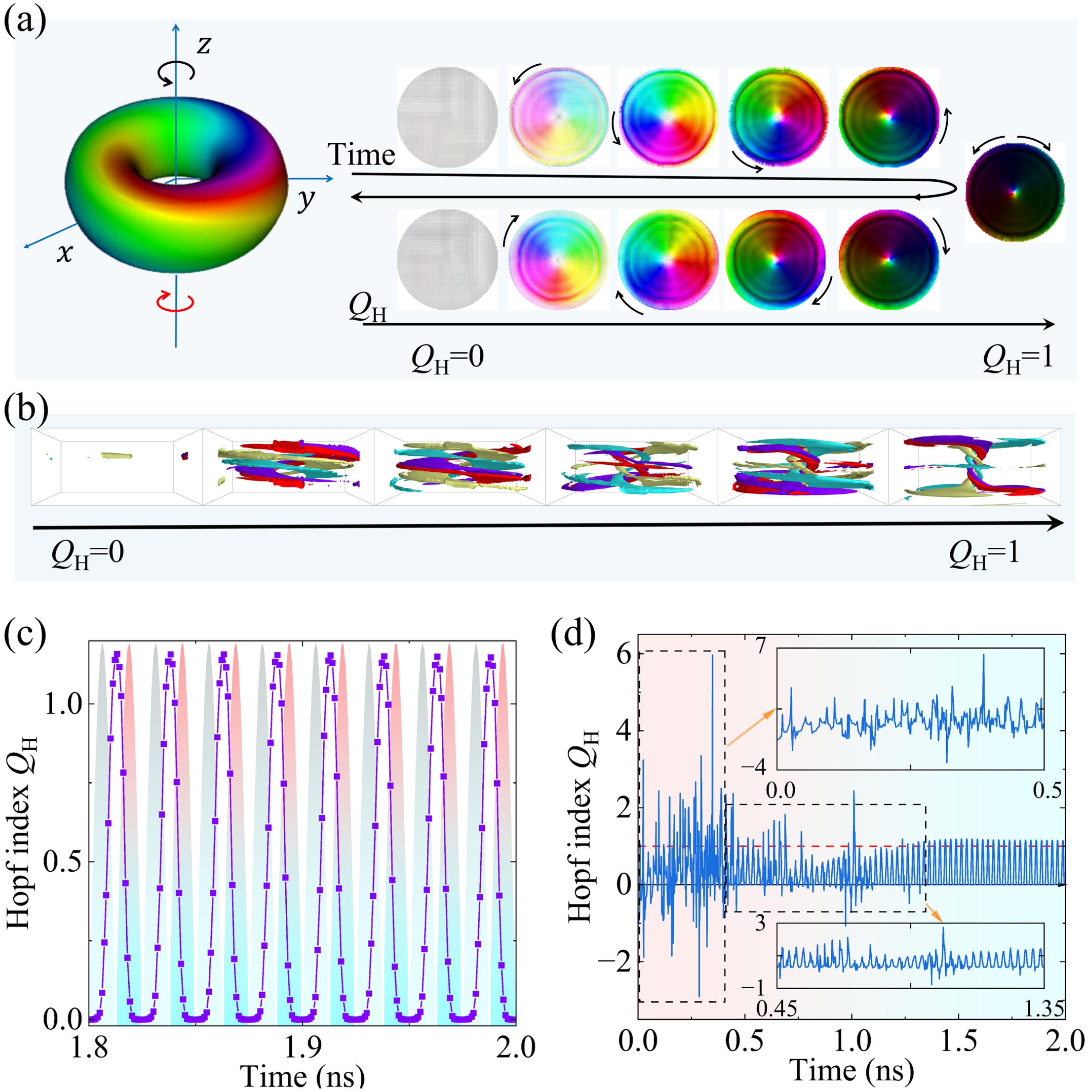}
\caption{(a) Temporal evolution of the magnetization texture during nonresonant rotational topological switching under a strong AM field with $B_0=10$ T and $f=40$ GHz. The frequency exceeds the intrinsic Hopfion-mode spectrum obtained in the weak-field regime, indicating that strong Zeeman torque and field locking, rather than linear eigenmode resonance, drive the rotational pathway. (b) Evolution of representative preimage links during the first half-cycle. (c) Hopf invariant $Q_{\rm{H}}$ as a function of time, showing reversible switching synchronized with the applied field. (d) Detailed time evolution of $Q_{\rm{H}}$, showing an initial transient regime, a metastable locking regime, and a final steady periodic switching state (inset).}
\label{fig-rotation}
\end{figure}

A qualitatively different pathway emerges as the driving frequency increases. Figure~\ref{fig-rotation}(a) shows the response to a strong AM field with $B_0=10$ T and $f=40$ GHz. In this case, the system undergoes rotational switching. Starting from an FM configuration, the spin texture rotates, develops the toroidal geometry of a Hopfion, and then returns to the FM state. This rotational motion is absent in the 2 GHz case, indicating that the drive's dynamical time scale determines the switching pathway; unlike the weak-field irreversible transition, the 40 GHz rotational switching does not arise from excitation of the linear Hopfion intrinsic mode. Instead, it occurs in a strongly nonlinear regime where the large Zeeman torque directly drives rapid precession of the transverse magnetization components and locks the texture's collective rotational motion to the external field (see S4 in the SM for details) \cite{ljing2022, slavin2009a}. The dimensionless driving strength $\chi=\gamma B_0/2\pi f$ is much larger than 1 for $B_0=10$ T and $f=40$ GHz, indicating that the field acts as a nonperturbative drive rather than a weak resonant perturbation. The resulting dynamics should therefore be viewed as a field-locked nonlinear limit cycle rather than the response of a single intrinsic internal mode \cite{ljing2022, jia2024}.

The preimage evolution in Figure~\ref{fig-rotation}(b) confirms that the rotational dynamics are genuinely topological: initially unlinked preimages become linked in the Hopfion phase and then unlink again as the system returns to the FM state. The Hopf invariant in Figure~\ref{fig-rotation}(c) oscillates periodically between the trivial and Hopfion states, synchronized with the AM field. The energy components, including the Zeeman energy, share this periodicity, as shown in Figure S7 of the SM \cite{supply}.

The route to steady switching differs from that of the breathing mode. Figure~\ref{fig-rotation}(d) shows three stages in the evolution of $Q_{\rm{H}}$. The system first undergoes a transient, nonsteady response, then enters a metastable regime where the dynamics begin to lock to the external drive. Finally, it reaches a stable periodic switching state. This locking process indicates that the rotational pathway requires a finite activation time before coherent topological switching is established.

Comparing the 2- and 40-GHz cases shows that strong AM fields can access dynamical regimes beyond the weak-field picture. At low frequencies, the texture can follow the drive via breathing-type expansion and contraction, which is connected to the low-frequency collective response of the confined Hopfion. At high frequency, the radial degree of freedom cannot follow the field adiabatically; instead, the large Zeeman torque drives rapid precession of the transverse spin components, producing a rotational, field-locked limit cycle. Thus, the field amplitude determines whether the topology-changing dynamics are irreversible or reversible, whereas the frequency selects which nonlinear collective coordinate becomes phase-locked to the drive. This separation of control parameters provides a general strategy for manipulating 3D topological solitons in confined magnetic systems. Here, reversibility means repeatedly recovering Hopfion configurations during continued periodic driving; it does not imply thermodynamic reversibility or retention of two independently writable states after the field is removed. The breathing and rotational examples show that periodic forcing can repeatedly reconstruct a three-dimensional magnetic texture through different collective motions. Their dependence on amplitude and frequency should nevertheless be understood jointly, since changing the frequency also changes the competition between forced precession, internal relaxation, and spatial deformation. The 10--11 T peak fields used here pose a substantial experimental challenge at GHz frequencies, and the results do not establish a lower bound on the field required in other geometries or materials. In the present geometry, we did not obtain complete reversible cycles at substantially lower amplitudes; lower drives produced only partial or non-reversible responses. We therefore regard these calculations as a theoretical demonstration of driven topological reconstruction. Establishing a practical operating regime will require systematically optimizing confinement, magnetic parameters, and the excitation protocol, along with robustness tests against thermal fluctuations. Our previous simulations show stable, reversible switching only within narrow amplitude-frequency windows and at high magnetic field strengths for 3D Hopfions, compared with 2D topological structures such as skyrmions. The magnetic field strengths at which the skyrmions undergo a similar reversible transition are approximately 4.1 T and 4.5 T \cite{fuTiltedChiralSpin2025}.

In summary, we have shown that AM fields drive two distinct dynamical topological regimes of confined magnetic Hopfions. Weak resonant fields can excite intrinsic modes and trigger a nonlinear instability that irreversibly converts a Hopfion into a toron state. By contrast, strong AM fields produce reversible GHz switching between topologically trivial FM configurations and Hopfion states. The switching pathway is frequency-selective: a 2-GHz drive induces breathing-mediated switching, whereas a 40-GHz drive induces rotational switching. Our results reveal how resonant and nonresonant nonlinear-driven dynamics govern the stability and controllability of 3D magnetic topology, and they establish the AM field amplitude and frequency as dual knobs for deterministic Hopfion manipulation.

\section*{ACKNOWLEDGMENTS}  This work is supported by Jiangsu Key Laboratory of Low-Dimensional Material Physics and Physical State Manipulation Open Projects Fund (No. LDMP2602), the Science and Technology Program of Xuzhou (KC25001) and by the National Natural Science Foundation of China (Grant Nos. 12374079, 11604380).


\bibliography{ref}

\end{document}